\documentclass[prd,tightenlines,showpacs,nofootinbib,eqsecnum,amsfonts,amssymb,amsmath,twocolumn,widetext]{revtex4-1}
\usepackage{graphicx}
\usepackage{amsmath}
\usepackage{amsfonts}   % if you want the fonts
\usepackage{amssymb}  
\usepackage{mathrsfs}
\usepackage{epsfig,bbm}
\usepackage{bm}
\usepackage{color}

\newcommand{\deu}{${\rm D}$}
\newcommand{\tro}{$^3{\rm He}$}
\newcommand{\qua}{$^4{\rm He}$}

\newcommand{\sep}{$^{7}{\rm Li}$}

\newcommand{\hli}{$^4$He, D, $^3$He and $^{7}$Li}

\newcommand{\gap}{\mathrel{ \rlap{\raise.5ex\hbox{$>$}}
                    {\lower.5ex\hbox{$\sim$}}  } }
\newcommand{\lap}{\mathrel{ \rlap{\raise.5ex\hbox{$<$}}
	            {\lower.5ex\hbox{$\sim$}}  } }

\newcommand{\etal}{{\it et al.}}
\newcommand{\dd}{{\rm d}}

\def\iso#1#2{\mbox{${}^{#2}{\rm #1}$}}
\def\h#1{\iso{H}{#1}}
\def\he#1{\iso{He}{#1}}
\def\li#1{\iso{Li}{#1}}
\def\be#1{\iso{Be}{#1}}

\begin{document}

\title{Mirror matter can alleviate the cosmological lithium problem}

\author{Alain Coc}
 \email{coc@csnsm.in2p3.fr}
 \affiliation{Centre de Spectrom\'etrie Nucl\'eaire et de
Spectrom\'etrie de Masse (CSNSM), IN2P3-CNRS and Universit\'e Paris Sud 11, UMR~8609, B\^at. 104, 91405 0rsay
Campus (France)}

\author{Jean-Philippe Uzan}
 \email{uzan@iap.fr}
 \affiliation{(1) Institut d'Astrophysique de Paris,
              UMR-7095 du CNRS, Universit\'e Pierre et Marie
              Curie,
              98 bis bd Arago, 75014 Paris (France),\\
              (2) Sorbonne Universit\'es, Institut Lagrange de Paris, 98 bis bd Arago, 75014 Paris (France).}

\author{Elisabeth Vangioni}
 \email{vangioni@iap.fr}
 \affiliation{(1) Institut d'Astrophysique de Paris,
              UMR-7095 du CNRS, Universit\'e Pierre et Marie
              Curie,
              98 bis bd Arago, 75014 Paris (France),\\
              (2) Sorbonne Universit\'es, Institut Lagrange de Paris, 98 bis bd Arago, 75014 Paris (France).}

\begin{abstract}
The abundance of lithium-7 confronts cosmology with a long lasting problem between the predictions of standard big bang nucleosynthesis and the baryonic density determined from the cosmic microwave background observations. This article investigates the influence of the existence of a mirror world, focusing on models in which neutrons can oscillate into mirror neutrons. Such a mechanism allows for an effective late time neutron injection, which induces an increase of the destruction of beryllium-7, due to an increase of the neutron capture, and then a decrease of the final lithium-7 abundance. Big bang nucleosynthesis sets constraints on the oscillation time between the two types of neutron and the possibility for such a mechanism to solve, or alleviate, the lithium problem is emphasized.
\end{abstract}
 \date{\today}
 \maketitle

%%%%%%%%%%%%%%%%%%%%%%%%%%%%%%%%%%%%%%%%%%
\section{Introduction}

The abundances of the light elements produced during the primordial nucleosynthesis (BBN) in the early hot phase of the Universe is one of the historical pillar of the big-bang model~\cite{jpubook}. The prediction of these abundances depends on:
\begin{enumerate}
\item the theory describing gravitation, which determines the cosmic expansion rate during BBN via the Friedmann equation,
\item the microphysics of the non-gravitational sector: properties of the particles and their interactions (mostly the weak interaction)
and nuclear physics (properties of nuclei and reaction cross--sections),
\item the matter content of the Universe, mostly through the number of baryons per photon $\eta=n_{\rm b}/n_{\gamma}$. 
\end{enumerate}
A key physical quantity is the freeze-out temperature of the weak interaction, $T_{\rm f}$, which can roughly be estimated from the Friedmann equation as
$$
G^2_{\rm F} T^5_{\rm f} \sim \sqrt{G g_*} T^2_{\rm f},
$$
so that we expect the predictions to depend on the number of relativistic degrees of freedom, $g_*$, the Fermi constant, $G_{\rm F}$, the Newton constant $G$ but also the nuclear quantities such as the lifetime of the neutron, the neutron to proton mass difference, as well as the fine structure constant.  
The subsequent nucleosynthesis is delayed because of the small deuterium binding energy and depends on the cross--sections of a dozen nuclear reactions.

As a consequence, BBN has been extensively used to constrain various derivations from the standard physical models such as scalar-tensor theories~\cite{dp,bbnus1,bbnus2}, variation of fundamental constants~\cite{jpucte,Coc07,cdouv,cte-other}, 
the existence of extra relativistic degrees of freedom, or the existence of particles decaying during BBN (see e.g. Ref.~\cite{keithTASI} for a review on BBN and physics beyond the standard model).  
We refer to Refs.~\cite{bbnrevue1,bbnrevue2} for reviews on the basics of BBN and on the constraints it sets on cosmology and physics.

Many of these extensions have been motivated by a long lasting problem of compatibility between \sep\ theoretically predicted abundances
and observations. 
When using for $\eta$ the value determined by the cosmic microwave observations, the BBN predictions for \qua, \deu\ and \tro\  are in very good  agreement
with those deduced from observations. 
However, there remains, a yet unexplained,  discrepancy of a factor $\approx$3, between  the calculated and observed \sep\ abundances,
that has not been reduced, neither by recent nuclear physics experiments, nor by new observations  
(see \S~\ref{s:obs} below for a detailed discussion of the lithium problem). 

There are two main channels to produce \sep. The first channel, $\he4+\h3\rightarrow\li7+\gamma$, is dominant for $\eta\lesssim 3.3\times10^{-10}$ while the second, $\he4+\he3\rightarrow\be7+\gamma$ followed by a $\beta$ decay with a half-life of 53~days, is dominant for $\eta\gtrsim 3.3\times10^{-10}$ (hence for WMAP value). 
A possibility to modify the abundance of \sep\ for the larger values of $\eta$ is to inject neutrons during the late stages of BBN \cite{Jed04,Coc07,Alb12}. 
That would reduce the amount of produced \be7, and thus of \sep, since it increases its destruction  due to a more efficient neutron capture.\\

Injecting neutrons is indeed not something easily performed without modifying the laws of nature and in particular the matter sector by including physics beyond the standard model of particle physics. Interestingly, such an idea can be realized by introducing a mirror sector. Such a sector is constructed by assuming that the gauge group $G$ of the matter sector  is doubled to the product $G\times G'$.  Imposing a mirror parity under the exchange $G\leftrightarrow G'$ implies that the Lagrangian of the two sectors, ordinary and mirror, are identical so that they have the same particles content such that ordinary (resp. mirror\footnote{Any field $\phi$, fundamental or composite, in the ordinary sector is associated to a field $\phi'$ in the mirror sector so that $n'$ corresponds to mirror neutrons, $p'$ to mirror protons etc.}) matter fields belonging to $G$ (resp. $G'$) are singlets of $G'$ (resp. $G$). They also have the same fundamental constants (gauge and Yukawa couplings, Higgs vev). The latter point implies that the microphysics (and in particular the nuclear sector) is identical in both sectors. The two sectors are coupled through gravity, and can eventually interact via some couplings so that the general form of the matter Lagrangian is
$$
{\cal L} = {\cal L}_G(e,u,d,\phi,\ldots) + {\cal L}_G(e',u',d',\phi',\ldots) +{\cal L}_{\rm mix}.
$$
Such a sector was initially proposed by Li and Yang~\cite{LiYang} in an attempt to restore global parity symmetry and was then widely investigated~\cite{mirror-gen,sterile}. Any neutral ordinary particle, fundamental or composite, can be coupled to its mirror partner hence leading to the possibility of oscillation between ordinary and mirror particles~\cite{mirror-osc}. For instance a mixing term of the form ${\cal L}_{\rm mix}\propto F_{\mu\nu}'F^{\mu\nu}$ will induce a photon-mirror photon oscillation, ordinary neutrinos can mixed with mirror neutrinos and oscillate in sterile neutrinos~\cite{sterile}. Among all the possible mixing terms, special attention has been drawn~\cite{mirrorneutron} to the mixing induced between neutrons and mirror neutrons. Such a possibility is open as soon as ${\cal L}_{\rm mix}$ contains a term $\propto (udd)(u'd'd') + (qqd)(q'q'd')$; see e.g. Ref.\cite{mirrorneutron} for details. It was also pointed out~\cite{mirrorneutron}  that a neutron--mirror neutron oscillation could be considerably faster than neutron decay, which would have interesting experimental and astrophysical implications.

This has motivated experimental searches for $nn'$-oscillations~\cite{mirrorlab1} which provide the constraint~\cite{mirrorlab2,PDG}
$$
 \tau_{\rm osc} >414\, {\rm s} 
$$
at 90\% C.L. on the oscillation time. Recently, it has been improved~\cite{exp-mirror-n1} to $\tau_{\rm osc} >448\, {\rm s}$ ( 90\% C.L).

From a cosmological point of view, mirror particles have been advocated as a dark matter candidate~\cite{mirrorDM,mirrorBBNearly}. In particular, mirror baryons do not interact with photons and have the same mass as ordinary baryons. From our world they can thus be considered as stable, self-interacting  dark matter particles. The interest in this candidate has been revived after it has been argued that mirror photons could provide an explanation of the direct detection experiments~\cite{DMnew}.

This demonstrates that such a sector is well-motivated from theoretical, experimental and cosmological points of view.

Coming back to our primary interest, namely BBN, the mirror electrons, photons and neutrinos act as extra-degrees of freedom. It is thus clear that it will modify the abundance of the light elements in the ordinary world and also that one needs to compute the abundances in the mirror world. In this framework, the abundances of the light elements are governed by 4 parameters $\lbrace\eta,\eta',\tau_{\rm osc},x \rbrace$, namely the two photon to baryon ratios, $\eta$ and $\eta'$,  the ratio between the photon temperature of the two worlds today, $x$, and the oscillation time $\tau_{\rm osc}$. Two main effects have to be taken into account: (1) the fact that the mirror sector accounts for extra-relativistic degrees of freedom, hence modifying the expansion rate during BBN, and (2) the oscillation between neutrons and mirror neutrons. Note again that the nuclear physics in the two worlds is not modified and strictly identical.\\

Early investigations have mostly relied on the first effect and studied the nucleosynthesis in the mirror world~\cite{mirrorBBNearly}. If the temperatures in both worlds are identical, then the effect of the mirror world is  equivalent to an effective number of neutrino families, $\delta N_{\rm eff}=6.14$~\cite{groom}, too large a  number to be compatible with observation. The goal of this article is thus two-fold. First, we want to investigate the effect of a mirror sector on the BBN predictions. In particular, we will show that it allows to set constraints on the parameter $\tau_{\rm osc}$ that are complementary to laboratory experiments. Then, secondly, within these limits, we shall show that there exists a range of parameters that alleviate, and eventually solve, the \sep\ problem.\\

The article is organized as follows. We start by recalling the observational landscape that sustains the lithium problem in \S~\ref{s:obs} and then describe the implementation of a mirror sector in BBN computation in \S~\ref{sec_bbn}. The effect of the mirror neutron migration is investigated in \S~\ref{sec_oscill} and the predictions are then compared with observations in \S~\ref{sec_obs}.

%%%%%%%%%%%%%%%%%%%%%%%%%%%%%%%%%%%%%%%%%%%%%%%%%%%%%
\section{The lithium problem} \label{s:obs}

Historically,  \sep\ obtained a cosmological status with the discovery of the Spite plateau~\cite{spitex2}. From an observational point of view, its abundance keeps a value constant over a large range of metallicity, (and effective surface temperature,) see e.g. Refs.~\cite{bm,rnb,Asplundetal06,Bonifacioetal07,hos,hos2,sbordone,ss4}, hence taken as the primordial value. 
From a theoretical point of view, as long as one sticks to standard BBN, the computation of the abundance of \sep\ requires only the knowledge of the baryon density, that is nowadays determined to an unprecedented precision from analyses of the CMB anisotropies and is given, based on the WMAP 7-year analysis~\cite{WMAP11} 
(see also the WMAP 9-year analysis~\cite{wmap9}) by, 
\begin{equation}
\Omega_{\mathrm{b}}{\cdot}h^2= \left(2.249^{+0.056}_{-0.057}\right) \,\times10^{-2}
\end{equation}
corresponding to:
\begin{equation}
\eta=\eta_{\rm WMAP}= \left(6.16^{+0.15}_{-0.16}\right) \times 10^{-10}.
\end{equation}
With this value, standard BBN gives an abundance  \sep/H of  $(5.24\pm 0.5) \times 10^{-10}$~\cite{CV10,coc12}, which is considerably higher than almost all observational determinations. The value determined in Ref.~\cite{Ryan00} is ${\rm Li/H}=(1.23^{+0.34}_{-0.16}) \times 10^{-10}$ and similarly, the recent analysis of  Ref.~\cite{sbordone} gave  ${\rm Li/H}  =(1.58 \pm 0.31) \times 10^{-10}$. This leads to  a clear discrepancy between theory and observation, the stellar observations leading to a value too low by a factor of 3--5. \\

%%%%%%%%%%%%%%%%%%%%%%%%%%%%%%%%%%%%%%%%%%%%%%%%%%%%%%%%
\begin{figure}[htb]
\begin{center}
\vskip -2.cm
 \includegraphics[width=.4\textwidth]{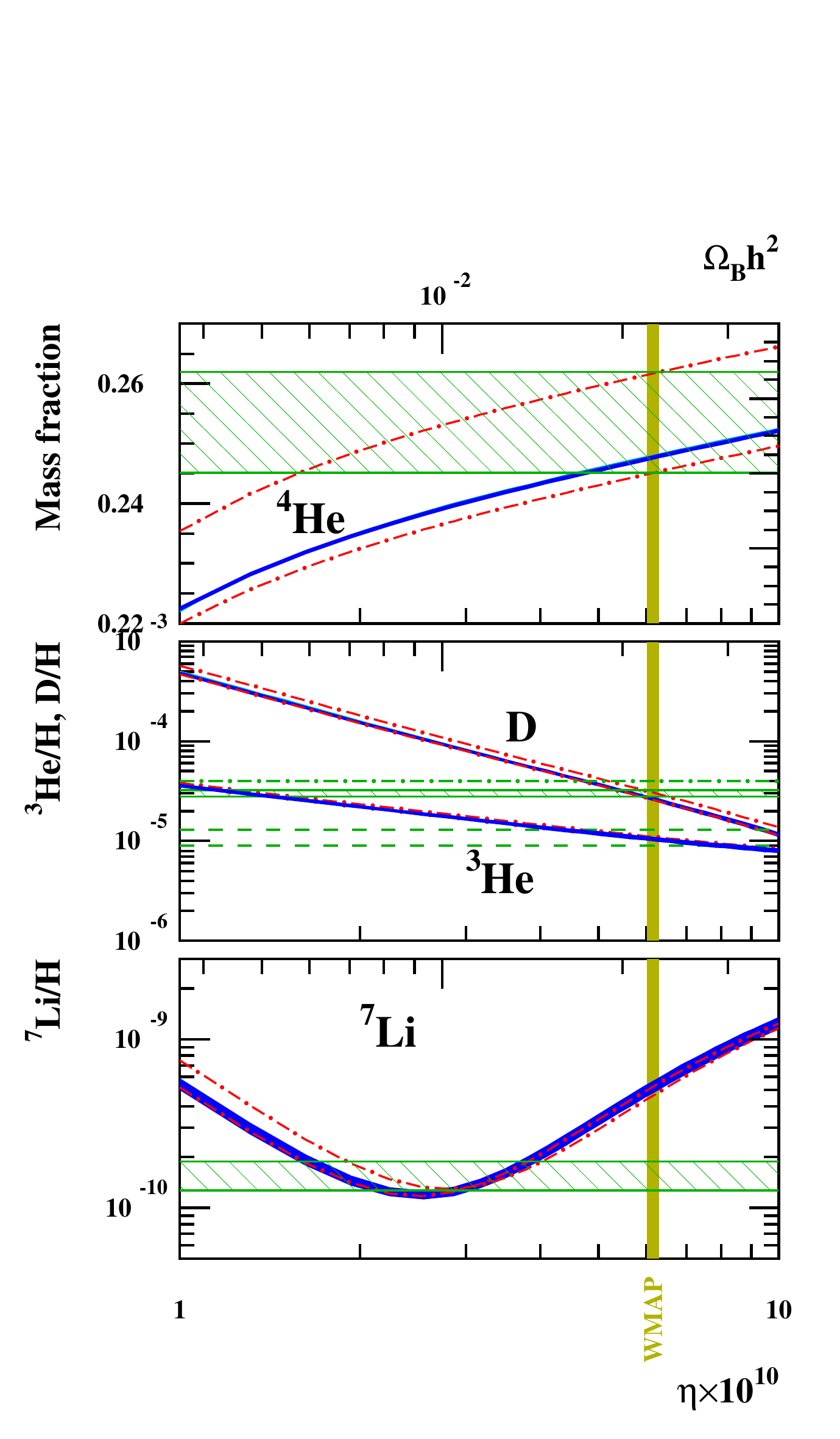}
%\begin{minipage}[b]{14pc}
\caption{\label{f:heli}
Abundances of \qua\, \deu, \tro\ and \sep\ as a function of the baryon over photon ratio 
$\eta$ (blue) \cite{CV10}. 
The vertical stripe corresponds to the WMAP baryonic density
while the horizontal area represent the adopted observational abundances (green region); see text.
The  dot-dashed lines correspond to the extreme values of the {\em effective} neutrino families
compatible with \qua\ observations, $N_{\rm eff}=(2.89,4.22)$.}
%\end{minipage}
\end{center}
\end{figure}
%%%%%%%%%%%%%%%%%%%%%%%%%%%%%%%%%%%%%%%%%%%%%%%%%%%%%%%%

Many studies have been devoted to the resolution of the so-called  \li7 problem and many possible ``solutions'', none of which fully satisfactory, have been proposed. 

First, the hypothesis underlying the standard computation and the stellar processes have all been reexamined carefully:  ($i$) the nuclear reaction rates used in BBN calculations have been reexamined~\cite{Cocetal04,angulo,cfo4,nuclear,boyd}; ($ii$) the possible depletion of \li7 in stars has been discussed~\cite{Ric05,Kornetal06,GarciaPerezetal08}; ($iii$) the temperature scale used in the \li7 abundance determination has also been the subject of a lively debate~\cite{hos2,mr2}; and ($iv$) the 3-dimensional stellar modeling and non-local thermodynamical equilibrium effects have also been investigated but the results are not very different from  those determined using 1-dimension models with local thermodynamical equilibrium \cite{Cay07}. 

Another possibility is that the solution of the \li7 problem could require new physics beyond the standard model.  Many models have been developed, among which ($i$) the decay of a massive particle during or after BBN that could affect the light element abundances and potentially lower the \li7 abundance~\cite{eov,Jedamzik06,pps,jp,grant3,jed08b,pp2,ceflos1.5,jittoh2,kk, Alb12,Cyb13,Jed04}. Recently, ($ii$) an axion condensate which cools the photon background leading to value of $\eta$ smaller than $\eta_{\rm WMAP}$ at the time of BBN  has also been advocated~\cite{sik,kus}.  Exotic solutions involving the possibility of a variation of the fundamental constants has also been debated~\cite{dfw,Coc07,cdouv}.  Note also the possibility of a modulation of the BBN conditions between, ($i$) the local Universe, where the stellar abundances are measured, and ($ii$) the large scale, on which the CMB is measured, so that the lithium problem can reflect a breakdown of the Copernican principle~\cite{cp1,cp2}. For a complete analysis, see the proceedings of the meeting ``Lithium in the cosmos''~\cite{LiinC}.\\

Despite this problem, standard BBN has been extremely successful and both  \qua\ and D/H show a good agreement between the theoretical predictions and their observational determinations (Fig.~\ref{f:heli}).

At $\eta = \eta_{\rm WMAP}$,  the \qua\ mass fraction is   $Y_p = 0.2476 \pm 0.0004$~\cite{CV10,coc12}. On the observational side, the determination of the helium abundance in extragalactic H{\sc ii} regions is somewhat difficult~\cite{os1}, Aver \etal~\cite{aos3} find a value with large uncertainties, $Y_p = 0.2534 \pm 0.0083$,  consistent with the BBN prediction. Using these observational limits, one can even set limits on the effective number of neutrino families (as defined in the next section) to be $2.89\leq N_{\rm eff}\leq4.22$ (red, dash-dotted curves in Fig.~\ref{f:heli}).

The observed deuterium abundance is also in a reasonable agreement with the BBN prediction which, for $\eta=\eta_{\rm WMAP}$,  is D/H = $(2.59 \pm 0.15) \times 10^{-5}$~\cite{CV10,coc12}. The weighted mean abundance of deuterium is D/H = $(3.02 \pm 0.23) \times 10^{-5}$ deduced from the observation of about 10  quasar absorption systems  (for more details see Ref.~\cite{Olive2012}).  Note that the individual measurements of D/H show considerable scatter and it is likely that systematic errors dominate the uncertainties. Moreover, the predicted abundance is somewhat {\em lower} than the observed mean one. Most of the measurements available in the literature have been reviewed in Ref.~\cite{pettini2} and more recently in Ref.~\cite{fuma}. However, the observation of a damped Lyman-$\alpha$ at $z_{\rm abs}=3.049$ has permitted~\cite{pettini12}  a new determination of the abundance of deuterium  ${\rm D/H} = (2.535 \pm 0.05) \times 10^{-5}$, leading to a mean determination lower than the previous one [$(2.60 \pm 0.12) \times 10^{-5}$]. However, since the H{\sc i} Ly-$\alpha$ absorption associated to this system is redshifted exactly on top of the Ly-$\alpha$-N{\sc v} blend emission from the quasar, the errors on this measurement are probably underestimated. Indeed the emission by the quasar cannot be recovered easily and is definitively degenerated with the exact profile of the absorption.
Taking into account this observation, Cyburt et al. \cite{Cyb13} find a mean value of  $(3.01\pm0.21)\times10^{- 5}$  which is comparable to the one taken in our
calculations.

However, it is important to note that the high dispersion of the observed D/H values at $z\sim 3$
(corresponding to $t$~=~3$-$4~Gyr) spans typically the $ (2.4 - 4.) \times 10^{-5}$  range. 
Different star formation histories in the galaxies associated with the DLAs can explain this dispersion.
D is a very fragile isotope (it is destroyed in stars at $T\sim10^5$~K) and
its destruction rate is dependent on the gas to total mass of a galaxy ratio
as shown in Ref.~\cite{vangioni88} (see their Figure 3). As a consequence, D/H destruction factors can range between 2 to  10. 
So, it is reasonable to consider that these pristine structures are in different evolution stages 
and consequently have different gas masses and, in situ, processing histories. 
Finally,  as D can only be destroyed in the course of chemical evolution, it is reasonable to relate the highest D/H 
observation to the primordial post-BBN abundance.
Accordingly, we consider that the primordial D/H abundance can be higher than the averaged value and
reasonably reach a value of $4.0 \times 10^{-5}$ (green, dash-dotted line in Fig.~\ref{f:heli}). 
For a recent analysis of the deuterium observations, we refer to Ref.~\cite{Olive2012}.

\section{Radiation era  and BBN with mirror matter}\label{sec_bbn}

\subsection{Evolution of the temperature of the mirror world}

\begin{figure}[htb]
\vskip 1cm
 \includegraphics[width=7cm]{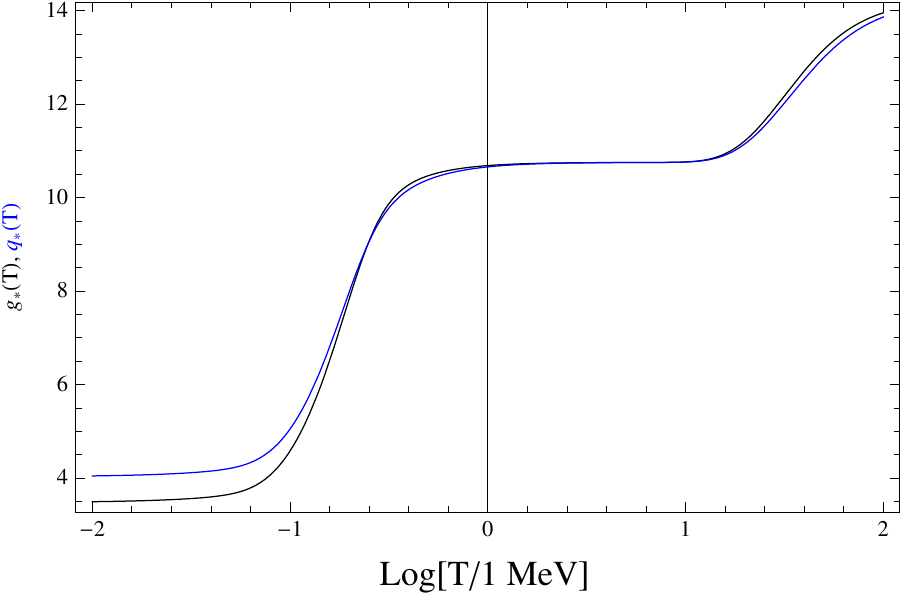}
 \caption{The evolution of $g_*$ and $q_*$ as a function of temperature $T$. At high temperature these two quantities are equal but differ after the neutrinos have decoupled. The second bumpy drop is due to the muon-antimuon annihilation at a temperature of order $m_e$.}
 \label{f:f1}
\end{figure}

The Friedmann equation contains both the ordinary and mirror matter so that during the radiation era it takes the form
\begin{equation}\label{e.fried}
 H^2 = \frac{8\pi G}{3}(\rho_{\rm r}+\rho'_{\rm r})
\end{equation}
where $G=M_{\rm p}^{-2}$ is the Newton constant, $\rho_{\rm r}$ the radiation energy density and where the contribution $\rho_{\rm r}'$ of the mirror sector has been included. $H=\dot a/a$ is the Hubble expansion rate of the scale factor $a$, a dot referring to a derivative with respect to the cosmic time. The energy density of the radiation can be expressed in terms of the temperature $T$ of the photon bath as
\begin{equation}
 \rho_{\rm r}(T) = \frac{\pi^2}{30}g_*(T) T^4
\end{equation}
where $g_*(T)$ is the effective number of relativistic degrees of freedom at the temperature $T$~\cite{jpubook}.  The determination of $g_*(T)$ follows the standard computation and has to include all relativistic particles. We recall that for any species of mass $m_i$, its energy density is given by
\begin{eqnarray}\label{e.rhoi}
 \rho_i &=& \frac{g_i}{2\pi^2}T^4\int_{z_i}^\infty \frac{\sqrt{u^2-z_i^2}}{\hbox{e}^u\pm1} u^2\dd u,
\end{eqnarray}
with $+$ for fermions and $-$ for bosons, and with $z_i=m_i/T$ and $u=E/T$. As depicted on Fig.~\ref{f:f1}, two effects are important at energy scales relevant for BBN: (1) the decoupling of the neutrinos after the freeze-out of the weak interaction, which implies that only photons are reheated during the electron-positron annihilation so that the temperature of the neutrinos $T_\nu$ differs from $T$. At temperatures smaller than the electron mass $m_e$ they are related by the relation
\begin{equation}
T_\nu = \left(\frac{4}{11}\right)^{1/3}T
\end{equation}
after the freeze-out , and (2) the particle-antiparticle annihilation (including electron-positron and muon-antimuon).  

%----------------------------------------------------
%----------------------------------------------------

Identical relations indeed hold for the mirror world. It is clear that if $T'_0=T_0$ today then $T'=T$ at all time so that the effect of the mirror world would be to double the numbers of relativistic of freedom, which would be in obvious disagreement with BBN. It follows that in order to be compatible with BBN observations, one needs $T'<T$ (i.e. $\rho_{\rm r}'<\rho_{\rm r}$ ) 
We thus set
\begin{equation}
T_0' = x T_0
\end{equation}
where $x$ is a new parameter of the model, which cannot be expressed in terms of previous ones. As long as $T,T'\ll m_e$ the temperatures in the two worlds have the same scaling with redshift so that $T'=xT$. However because the electron-positron annihilation and the decoupling of the weak interaction do not happen simultaneously in the two worlds, we expect $T'\not=x T$ in general. We thus set
\begin{equation}\label{alphaXT}
T'[x,T]=\alpha_x(T)xT,
\end{equation} 
with $\alpha_x(T)=1$ for $T\ll m_e$. 

\begin{figure}[htb]
 \includegraphics[width=7cm]{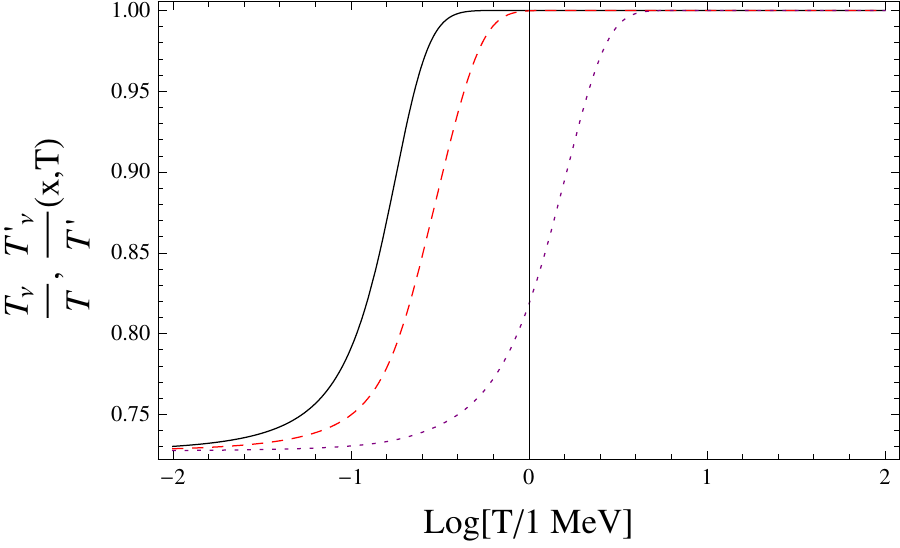}
 \caption{$T'_\nu/T'$ as a function of $T$ to $T_\nu/T$ for $x=1.$ (black, solid) $x=0.5$ (red, dashed) and $x=0.1$ (purple, dotted).}
 \label{f:f1a}
\end{figure}

Note that because the neutrinos are not reheated during the electron-positron annihilation, their temperature always scales as $(1+z)$ in both worlds so that
\begin{equation}\label{Tnuevo}
T'_\nu=xT_\nu
\end{equation} 
at all times. The relation between $T'$ and $T$ is thus obtained by solving 
\begin{equation}
 T'_\nu(T')=xT_\nu(T).
\end{equation}  
Figure~\ref{f:f1a} compares the evolution of $T'_\nu/T'$ as a function of $T$ to $T_\nu/T$ for $x<1$. Since the decoupling occurs before in the mirror world, the curves are shifted to higher $T$. If plotted as the function of $T_\nu$ instead of $T$, the curves will be identical up to a translation. 

%----------------------------------------------------
%----------------------------------------------------

\begin{figure}[htb]
 \includegraphics[width=7cm]{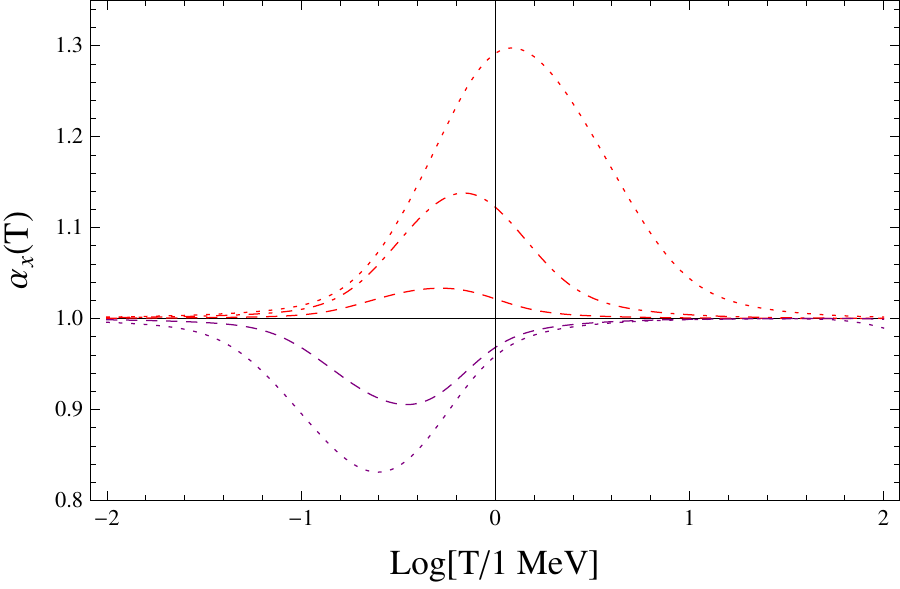}
 \caption{The evolution of the coefficient $\alpha_x(T)$, defined in Eq.~(\ref{alphaXT}) for $x=1$ (black), $x=0.8, 0.4, 0.1$ (red dashed, dot-dashed, dotted) and $x=1.2,2$ (purple, dashed, dotted). As discussed above, $\alpha_x(T)=1$ at low and high temperatures.}
 \label{f:f2}
\end{figure}

This translates to the evolution of $\alpha_x(T)$ as shown on Fig.~\ref{f:f2}. We recover the limiting behaviours that we have discussed above. The deviation from $\alpha_x=1$ occurs when one of the two temperatures is of order of $m_e$ and reaches a typical magnitude of (20-30)\%. The amplitude increases for smaller $x$ but then the contribution to the energy density is suppressed by a factor $x^4$ so that the effect on the Friedmann equation becomes negligible.

\bigskip

%----------------------------------------------------
%----------------------------------------------------

To finish, note however, and contrary to what has been assumed e.g. in Ref.~\cite{mirrorBBNearly} that even though at temperatures relevant for BBN, the ordinary and mirror neutrons are non relativistic and do not interact with photons and neutrinos, the entropy, $s=\frac{2\pi^2}{45}q_*T^3$, of  the two worlds are separately conserved but that $s'/s$ does not remain constant. Indeed,
\begin{equation}\label{TdeT}
 \frac{s'}{s} = \frac{q_*(T')}{q_*(T)}\alpha_x^3(T) x^3.
\end{equation}

\subsection{Cosmic dynamics in the ordinary world}

The Friedmann equation~(\ref{e.fried}) takes the form
\begin{equation}\label{e.fried2}
 H^2 = \frac{8\pi^3}{90} g_*(T)\left[1+K(T,T')x^4\right]\left(\frac{T^2}{M_{\rm p}}\right)^2,
\end{equation}
where
\begin{equation}\label{KTT}
 K(T,T') \equiv\frac{\rho'_{\rm r}(T')}{\rho_{\rm r}(T)}x^{-4}= \frac{g_*(T')}{g_*(T)}\alpha_x^4(T)
\end{equation}
is a factor that takes into account that a priori $T'\not=T$. The mirror sector appears as extra relativistic degrees of freedom. Seen from the ordinary world, this amount of extra-radiation is characterized by $K_x(T) \equiv K[T,T'(x,T)]$. The deviation from the standard Friedmann equation in the radiation era is often parameterized in terms of an effective number of neutrino species, $N_{\rm eff}$ as
\begin{equation}
 \rho_{\rm r}= \left[1+\frac{7}{8}N_{\rm eff}\left(\frac{T_\nu}{T}\right)^4 + \frac{15}{\pi^2}\frac{\rho_e}{T^4}
 \right]\rho_\gamma.
\end{equation}
from which $\delta N_{\rm eff}$ is  constrained to $2.89\leq N_{\rm eff}\leq4.22$ (\S~\ref{s:obs}). 
As depicted on Fig.~\ref{f:f3} (left), this implies that the effective number of neutrinos is temperature dependent and given by
\begin{equation}
 \delta N_{\rm eff} = \frac{4}{7}\frac{g_*(T)}{(T_\nu/T)^4} K_x(T) x^4.
\end{equation}
At low temperature, it reduces to $\delta N_{\rm eff} \simeq7.39K x^4$, while at high temperature it gives $\delta N_{\rm eff} \simeq6.14K x^4$, as used in Ref.~\cite{mirrorBBNearly}. For $T\ll m_e$, it reduces to the standard parameterization
$$
 \rho_{\rm r}= \left[1+\frac{7}{8}\left(\frac{4}{11} \right)^{4/3}N_{\rm eff}\right]\rho_\gamma
$$\\
from which the analysis of the WMAP data combined with baryonic acoustic oscillations sets~\cite{WMAP11} $2.8<N_{\rm eff}<5.9$ at 95\% C.L., compatible with the standard prediction $N_{\rm eff}\simeq3.046$.

\begin{widetext}
It is clear that if $T'=T$ then, because the two matter sectors are identical, $x=1$ and $K(T,T')=1$ so that $\delta N_{\rm eff}\sim7.39$ or $\delta N_{\rm eff}\sim6.14$ at low and high temperatures respectively. This is by far incompatible with the existing constraints mentioned above. As long as $x$ is too large, the effective radiation content of the Universe is also too large and BBN will not be successful. As seen from Fig.~\ref{f:f3} (right), we need typically that $x\lesssim0.6$ in order to have a chance to be compatible with current observations.
%----------------------------------------------------
\begin{figure}[htb]
\includegraphics[width=8cm]{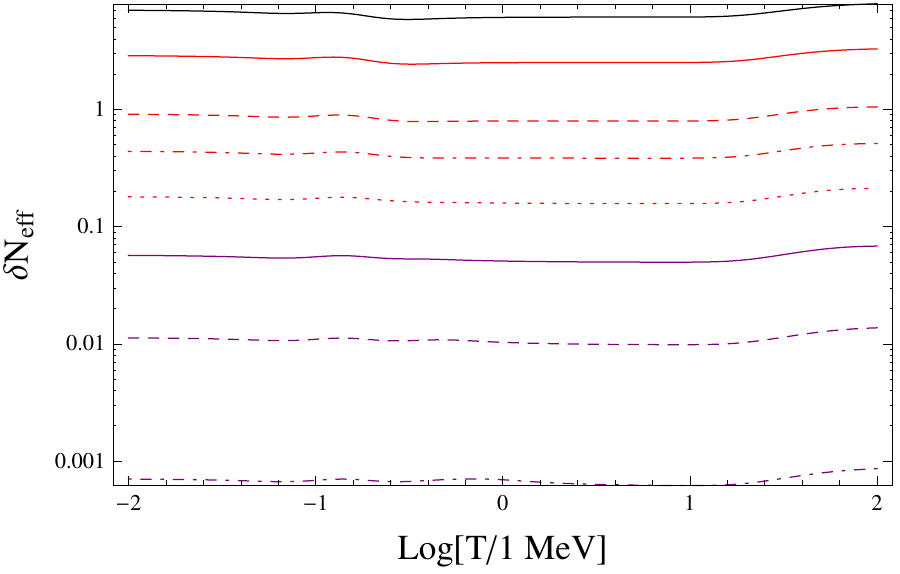} \includegraphics[width=8cm]{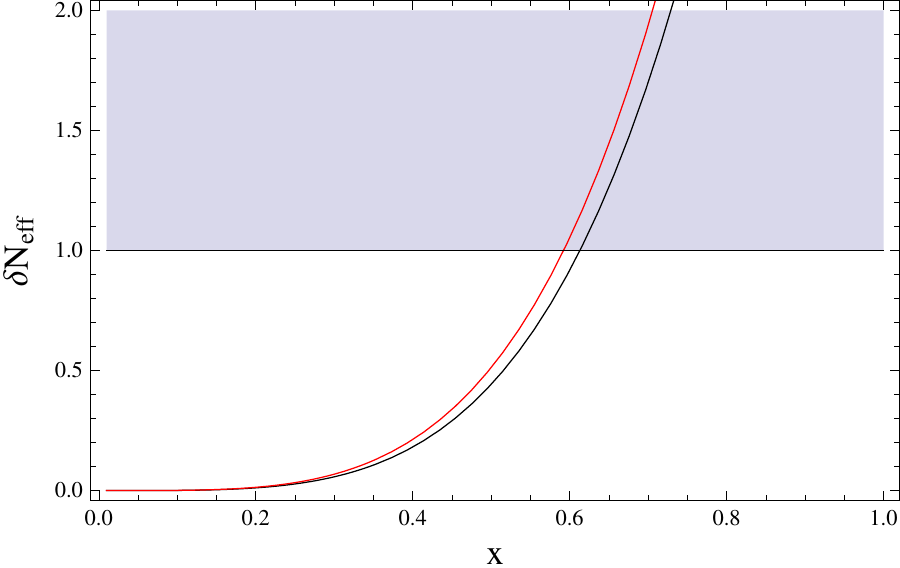} 
 \caption{Evolution of  $\delta N_{\rm eff}$ (left) with temperature for $x=$1 (black), $0.8, 0.6, 0.5,0.4$ (red), $0.3,0.2,0.1$ (purple).  The bump corresponds to the electron-positron annihilation while  the small increase at thigh temperature is due to the muon-antimuon annihilation. Evolution of  $\delta N_{\rm eff}$ (right) with $x$ for $T\ll m_e$ (black) and $m_e\ll T\ll m_\mu$ (red). The shaded region is experimentally excluded, so that we expect $x\lesssim0.6$.}
 \label{f:f3}
\end{figure}
%----------------------------------------------------
\end{widetext}

\subsection{Implementation of the mirror sector for BBN computation}

As explained in the introduction, the BBN computations depend on 4 parameters. The standard parameter is the number of baryons per photon, $\eta$. It has to be complemented by the constant ratio $x$, defined in Eq.~(\ref{TdeT}) and the number of baryons per photon in the mirror sector, $\eta'$, defined by  
\begin{equation}\label{eq:dens}
 \rho_{\rm b}'=\rho_{\rm b}{{\eta'}\over{\eta}}x^3
\end{equation}
where the factor $x^3$ accounts for the difference in photon number density. 

Since the physics in the two sectors is identical,  the reaction rates are the same in both worlds and BBN is the same except for the temperature and the baryonic densities. In order to allow for neutron oscillations, we include in the nuclear network normal and mirror isotopes with initial $n'$ and $p'$ abundances given by:
\begin{equation}
X_{\rm n',p'}=X_{\rm n,p}{{\eta'}\over{\eta}}x^3.
\label{eq:ratio}
\end{equation}
Accordingly, the mass fractions are such that
\begin{equation}
\sum_i X_i = 1+{{\eta'}\over{\eta}}x^3
\end{equation}
is conserved.

As discussed in the previous section, instead of the approximation $T'=xT$ for the temperature in the mirror sector and $\delta N_{\rm eff}\equiv6.14x^4$
to account for the accelerated expansion, we followed mirror electron and positron annihilations to calculate $T'$  [Eq.~(\ref{alphaXT})] and added the corresponding mirror radiation 
density, $\rho'$, in the Friedmann equation [Eqs.~(\ref{e.fried}) and (\ref{e.fried2})]. This does not change the qualitative global features of the results, as compared to previous
analyses, but leads to significant quantitative effects.

The network includes 16 isotopes and 27 reactions. These are the usual 8 isotopes and their mirror
partners, the 13 main BBN reactions and their mirror counterparts plus the $n\leftrightarrow n'$
oscillation term, described in the next section. The reaction rates are the same for ordinary and mirror reactions and are those 
used in previous works e.g. Ref.~\cite{CV10}, i.e. mostly from Ref.~\cite{Des04}, except
for  $^1$H(n,$\gamma$)D \cite{And06} and $^3$He($\alpha,\gamma)^7$Be \cite{Cyb08a}.

%----------------------------------------------------
\begin{figure}[htb]
%/Users/acoc/phys/ez_new/new_bbn/bbn_miroir/mirror_kumac/time_mirror.kumac
 \includegraphics[width=8cm]{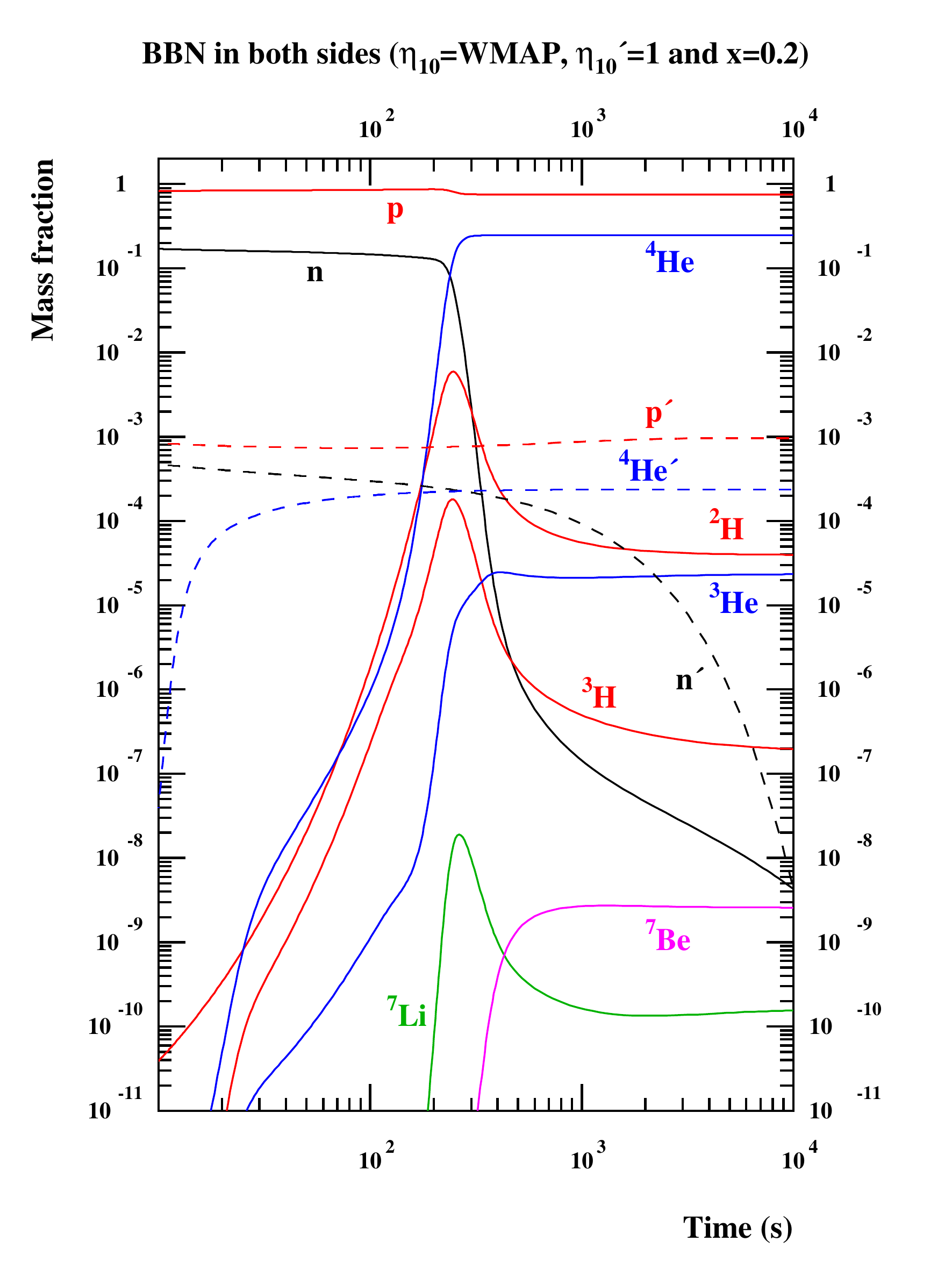} 
 \caption{Standard BBN (solid lines) and mirror BBN (dashed lines, only $n'$, $p'$ and \qua' are depicted, respectively in black, red and blue). This computation assumes that the two worlds do not communicate, i.e. that $\tau_{\rm osc}=\infty$.}
 \label{f:time0}
\end{figure}
%----------------------------------------------------

As an example, Fig.~\ref{f:time0} shows the result of a BBN calculation with $\eta=\eta_{\rm WMAP}$, $\eta'=10^{-10}$ and $x=0.2$. Nucleosynthesis proceeds in both worlds and the abundances in the mirror world are consistent with the early estimations of Ref.~\cite{mirrorBBNearly} (e.g. their Fig.~3). Since they are of no interest in this study (they do not modify the ordinary abundances and are not observable), the abundances of mirror isotopes other than  $n'$, $p'$ and \qua' are calculated but not displayed. 

The abundances of the mirror isotopes are globally reduced [Eq.~(\ref{eq:ratio})] with respect to the normal ones. However, it can be seen that after a time of order 300~s, the abundance of the mirror neutrons, $n'$, remains much higher than the normal neutron abundance. This is due to the fact that, for this choice of parameters ($T'$ and $\rho_{\rm b}'$), mirror BBN is limited to $n'$ decay while normal BBN is in its full development.  It is well known~\cite{Alb12,Jed04,Coc07} that  late time neutron injection  in BBN would alleviate the lithium problem by $^7$Be destruction through the  $^7$Be(n,p)$^7$Li(p,$\alpha)^4$He chain.  At WMAP baryonic density, \sep\ is produced indirectly by $^3$He($\alpha,\gamma)^7$Be, that will decay much later to \sep. This could be achieved, if some fraction of the mirror neutrons would migrate out of the mirror world into ours, as we shall now investigate.    

%%%%%%%%%%%%%%%%%%%%%%%%%%%%%%%%%%%%%%%%%%%%%%%%%%%%%%%%%
\section{Neutron oscillations}\label{sec_oscill}

In standard BBN, the weak interaction  maintains the thermal equilibrium between neutrons and protons until their rates become slower than the Hubble expansion rate at a typical temperature of $T\approx$~3.3~GK~\cite{ENAS6}. This freeze-out of the weak interaction is followed by neutron free decay until $T\approx$~0.9~GK when nucleosynthesis begins. This is during this phase of free decay that a mirror world allows for neutron oscillation between the two worlds. This interval of temperatures would be even wider in the mirror world if, as considered here,  $T'<T$ and $\rho_{\rm b}'<\rho_{\rm b}$ (see Fig.~\ref{f:time0}).

To introduce neutron oscillations, it is assumed that the mass matrix  is given in the form
\begin{equation}
M=\left(\begin{array}{cc}
m-i/2\tau & 1/\tau_{\rm osc} \\
1/\tau_{\rm osc} & m-i/2\tau\\
\end{array}\right),
\end{equation}
where $\tau$ is the neutron lifetime. It follows that the mirror neutron abundance evolves as 
$e^{-t/\tau}\cos^2(t/\tau_{\rm osc})$~\cite{Moh80} so that
\begin{equation}\label{eq:oscil}
\lambda_{{\rm n'}\rightarrow{\rm n}}=
- \frac{\rm d}{{\rm d}t}\log\cos^2\left(\frac{t}{\tau_{\rm osc}}\right)={2\over{\tau_{\rm osc}}}\tan\left(\frac{t}{\tau_{\rm osc}}\right).
\end{equation}
This rate looks like a decay rate $\tau_{\rm osc}/2$ but modulated by $\tan\left(t/\tau_{\rm osc}\right)$.  It diverges for $t/\tau_{\rm osc}=\pi/2$ but (putting aside numerical details), keeps the product  $X_{\rm n'}(t)\tan\left(t/\tau_{\rm osc}\right)$ finite, i.e.
\begin{equation}
\lim_{t/\tau_{\rm osc}\rightarrow\pi/2}X_{\rm n'}(t)\tan\left(t/\tau_{\rm osc}\right)=0 
\end{equation}
if there is no other source of mirror neutrons. This is indeed the case during free mirror neutron decay as we neglect $\lambda_{{\rm n}\rightarrow{\rm n'}}$. When  $X_{\rm n}(t){\gg}X_{\rm n'}(t)$, at small $t$, $\lambda_{{\rm n}\rightarrow{\rm n}'}$ is suppressed by a factor $\tan\left(t/\tau_{\rm osc}\right)$. Later, when this factor is no longer negligible, $X_{\rm n}(t){\ll}X_{\rm n'}(t)$
so that we can neglect neutron diffusion to the mirror sector (but not the reverse). 

%----------------------------------------------------
\begin{figure}[hbt]
%/Users/acoc/phys/ez_new/new_bbn/bbn_miroir/mirror_kumac/time_mirror.kumac
 \includegraphics[width=8cm]{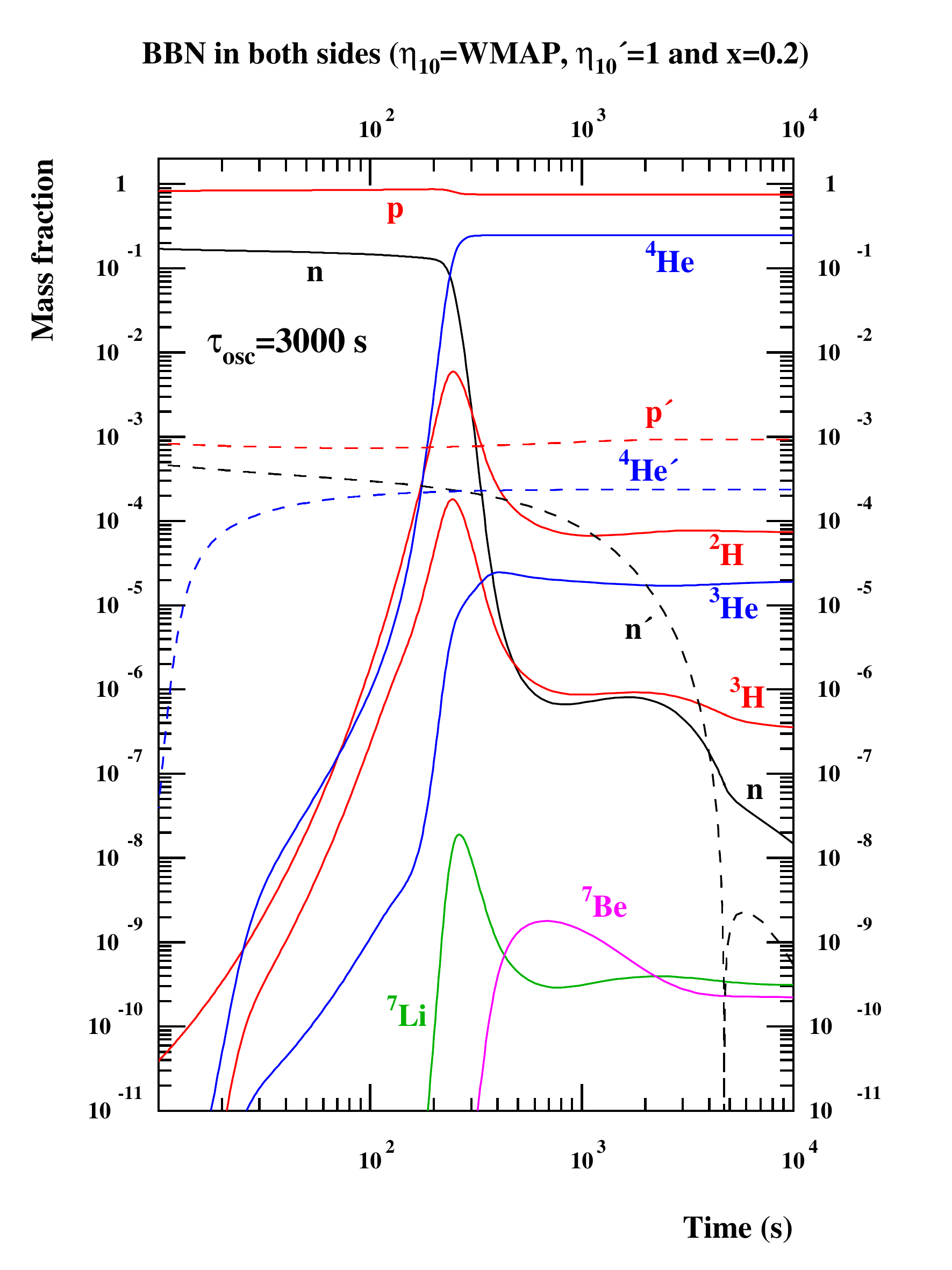}
 \caption{Same as Fig.~\ref{f:time0} but allowing neutron oscillations with $\tau_{\rm osc}=3\times10^3$~s: the $^7$Be+\sep\ abundance is 
 significantly reduced.}
 \label{f:time3}
\end{figure}
%----------------------------------------------------

Fig.~\ref{f:time3} shows the result of a calculation with the same parameters as in Fig.~\ref{f:time0} except that we assume oscillations 
from $n'$ to $n$ with $\tau_{\rm osc}=3\times10^3$~s in Eq.~(\ref{eq:oscil}), a value well above the experimental lower limit.   
Initially, because of the difference in baryonic densities, the neutron abundance 
is much higher in our world but, during standard BBN the neutron abundance decreases very rapidly compared to mirror BBN because of the higher temperature and density. Then,  when the $n$ abundance dropped below the $n'$ one, $n$ injection from oscillating $n'$, at the time of $^7$Be formation, leads to its destruction due to a more efficient neutron capture. On the contrary, the abundance of \deu\ increases: we obtain a reduction of the $^7$Li~+~$^7$Be abundance at the expense of a moderate higher deuterium.   We can observe a fast decrease of the mirror neutron  abundance, down to $\approx0$, around $t\approx\tau_{\rm osc}$, followed by a small increase.  This late time effect is due to the nucleosynthetic $n'$ production by the $^2$H'(d',p')$^3$H'(d',n')$^4$He' chain when the oscillation is turned off. This should not affect the results because it happens when the  $^7$Li~+~$^7$Be abundance has reached a plateau and because, in any case, these are {\em fresh}
mirror neutrons for which the oscillation clock has to be reset.

Note that, before freeze--out the weak interactions, that equilibrate the neutron and proton abundances, prevent any oscillation. We have assumed here that after freeze-out, and until p'(n',$\gamma$')d' reaction becomes efficient, neutrons are free to decay and oscillate.

\section{Compatibility with observations}\label{sec_obs}

\begin{widetext}
In the previous section, we have shown that the \sep\ primordial abundance can be reduced significantly for values of the parameters $x$, $\eta'$ and $\tau_{\rm osc}$ that were found to be representative of the favorable solutions. We now explore this parameters space and compare the predictions to the BBN for \hli\ primordial lithium (see \S~\ref{s:obs}). 
%----------------------------------------------------
\begin{figure}[htb]
%/Users/acoc/paw/BBN/mirror/2D/heli_miroir-ex.nb
 \includegraphics[width=5cm]{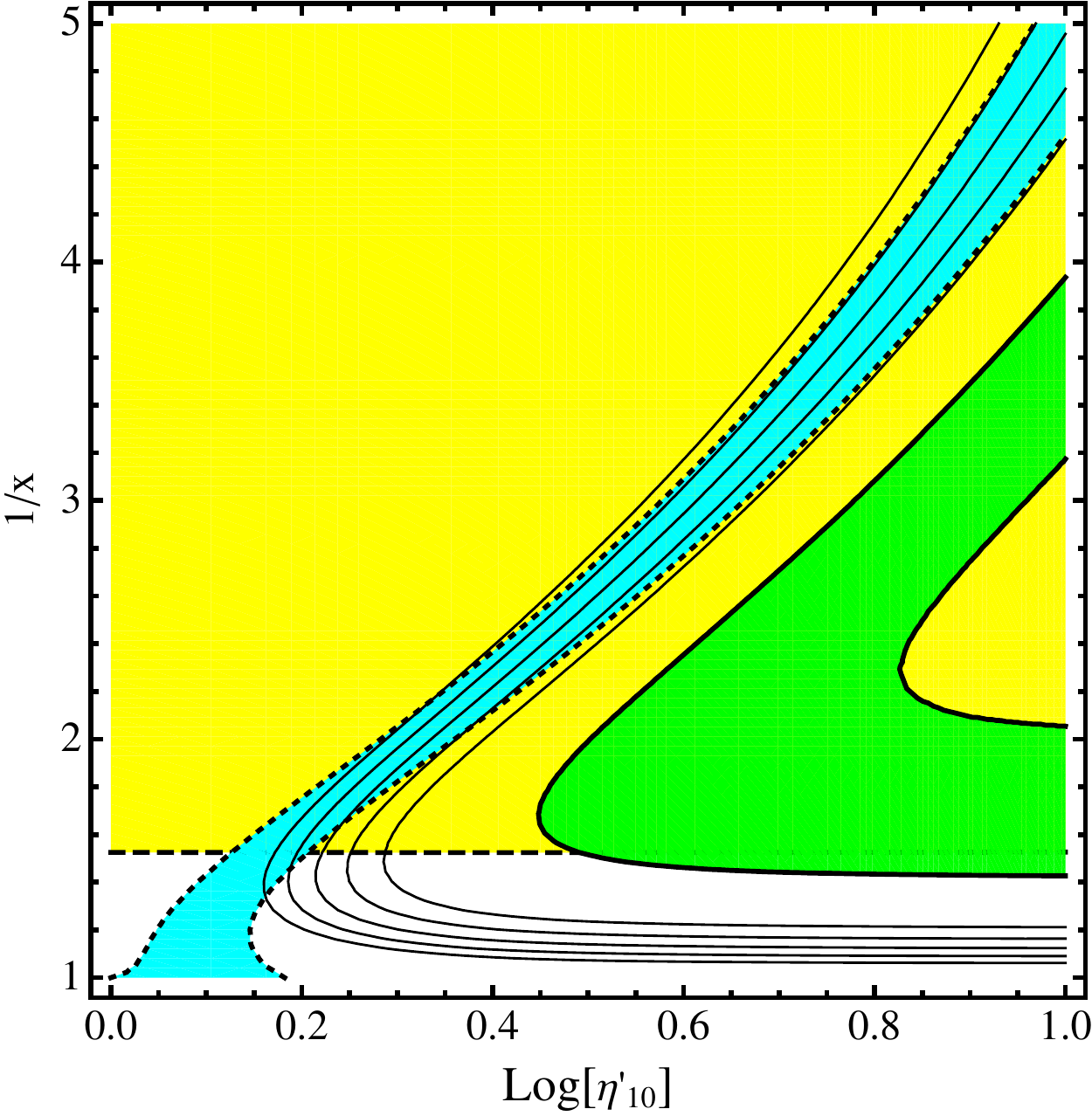}$\qquad$ \includegraphics[width=5.2cm]{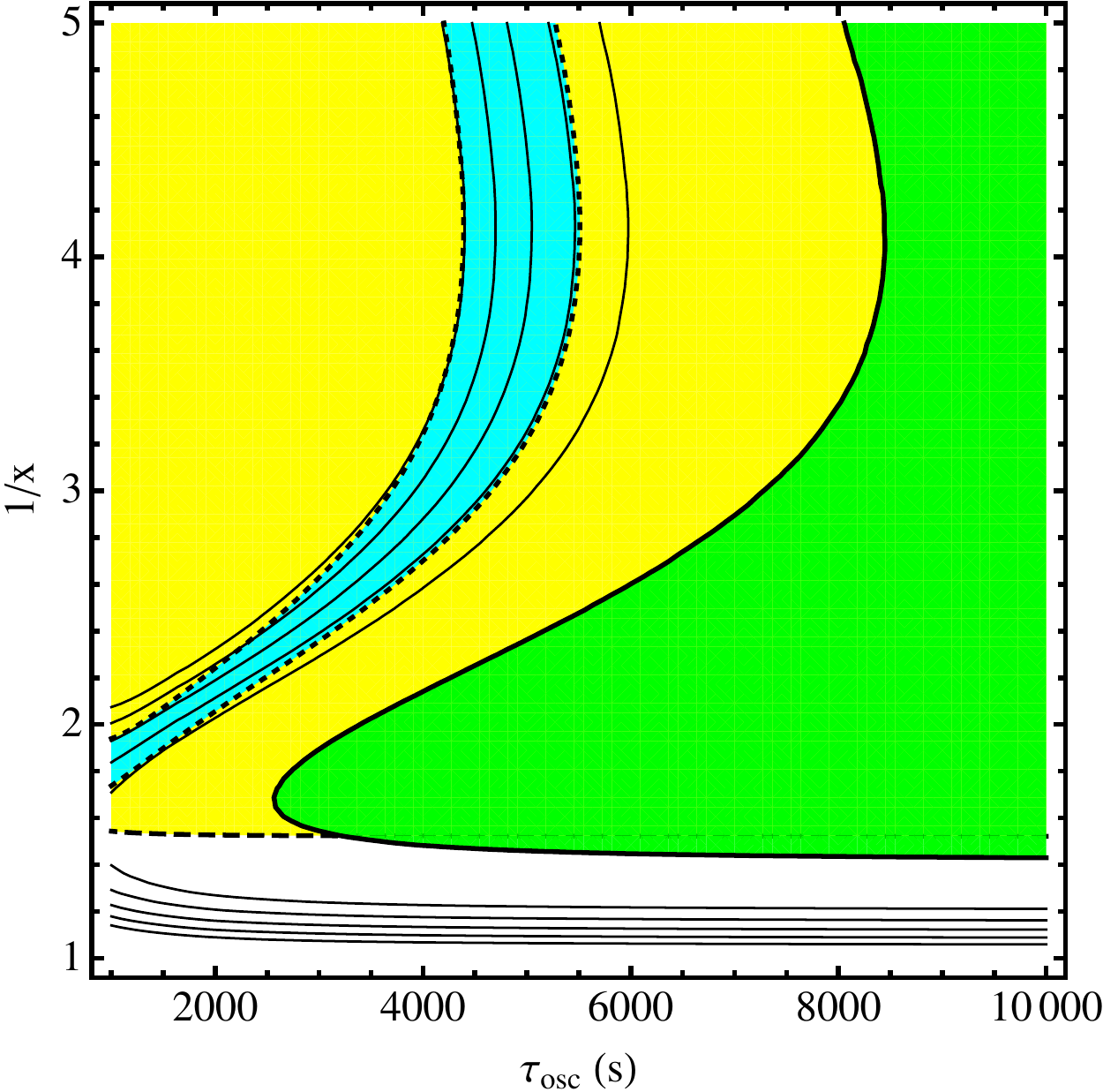}$\quad$  \includegraphics[width=5.4cm]{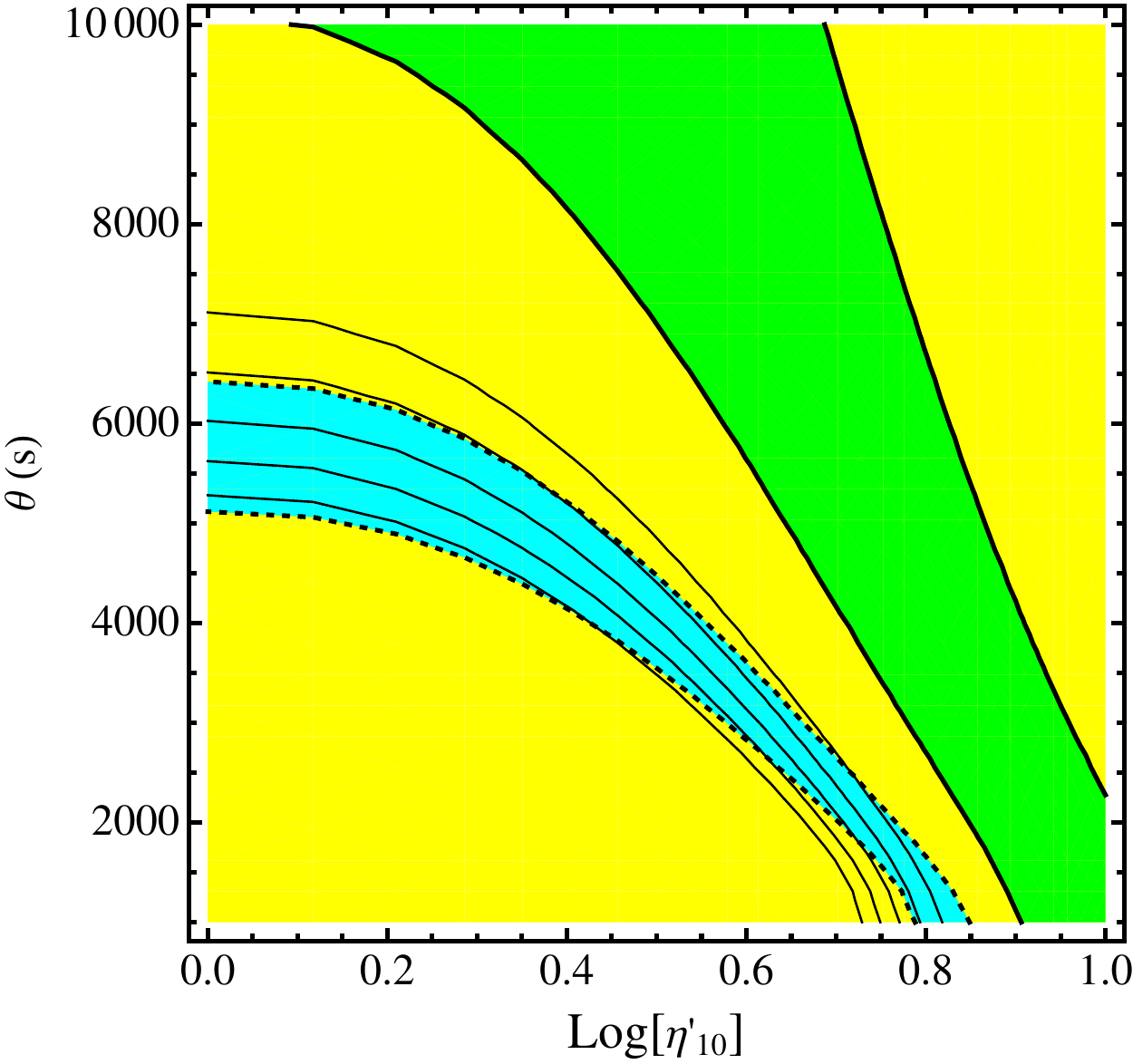}
 \caption{(left) Limits on $\eta_{10}'$ and $1/x$  provided by observational constraints  on D (solid; the green is acceptable, i.e. $2.79\times10^{-5}<{\rm D/H}<3.25\times10^{-5}$) \qua\ (dashed; the yellow region is acceptable, i.e. $Y_p=0.2534\pm0.0083$) and \sep\ (dotted; the light blue strip is acceptable, i.e. ${\rm Li/H}=(1.58\pm0.31)\times10^{-10}$),  allowing for neutron oscillations with $\tau_{\rm osc} = 3\times10^3$~s constant. The parameters space in the blue strip reconciled both \sep\ and \qua, but not \deu. This conclusion is changed if one allows for a larger value of D/H. The thin black lines corresponds, from right to left, to D/H = $(3.8, 4.0, 4.2, 4.4, 4.6)\times10^{-5}$, which shows that for D/H $\in [3.8, 4.6]\times10^{-5}$ there exists a region of the parameters $(x,\eta'_{10})$ reconciling \sep\ and \qua\ and \deu.
 (middle) Same but as a function of  $\tau_{\rm osc}$ and $1/x$, keeping $\eta_{10}'=3$ constant. 
 (right) Same but as a function of  $\eta_{10}'$ and $\tau_{\rm osc}$, keeping $1/x=3$ constant. }
  \label{f:heli-ex}
\end{figure}
%----------------------------------------------------
\end{widetext}

In Fig.~\ref{f:heli-ex} (left),  we first explore the parameters space $10^{-10}<\eta'<10^{-9}$ and $0.2<x<1$ for a fixed value of $\tau_{\rm osc}= 3\times10^3$~s, assuming neutron oscillations as in Eq.~\ref{eq:oscil}. It shows that for a wide range of parameters, \sep\ and \qua\ abundances are compatible with observations (overlap between the blue and yellow areas), but at the expense of \deu\ (the green area). [The constant 1/$x$ lower limit given by \qua\ just reflects the $\delta N_{\rm eff}\leq1.22$ limit (from \S~\ref{s:obs})].   
However, it also shows that, if we rather adopt a primordial \deu/H abundance of $\approx$4$\times10^{-5}$, which is reasonable as discussed in \S~\ref{s:obs}, there exists a region in the parameter space $(x,\eta')$ for which one obtains a full compatibility for all, \qua, \deu\ and \sep\ isotopes. This is confirmed by the Figs.~\ref{f:heli-ex} (middle and right) leading to the conclusion that there exists a region in the parameters space $(x,\eta',\tau_{\rm osc})$ for which the lithium-7 problem is solved, at the expense of a 
moderately higher D/H.
This is, in fact, a common feature that models succeeding in reducing the \sep\ BBN production come with an over production of D.  This nevertheless can be cured e.g. by non-thermal photons injection by the radiative decay of long-lived particles after the BBN \cite{kus}.
Fig.~\ref{f:heli-ex} (right), shows that  if we adopt the mean value observed today, ${\rm D/H} = (3.02 \pm 0.23) \times 10^{-5}$, for any $\tau_{\rm osc}$ values  
between  $10^3$ and $10^4$~s there are combinations of  $\eta'$ and $x$,  that reconcile D/H and helium-4 (the green area).

\begin{figure}[hbt]
%/Users/acoc/phys/ez_new/new_bbn/bbn_miroir/mirror_kumac/time_mirror.kumac
 \includegraphics[width=8cm]{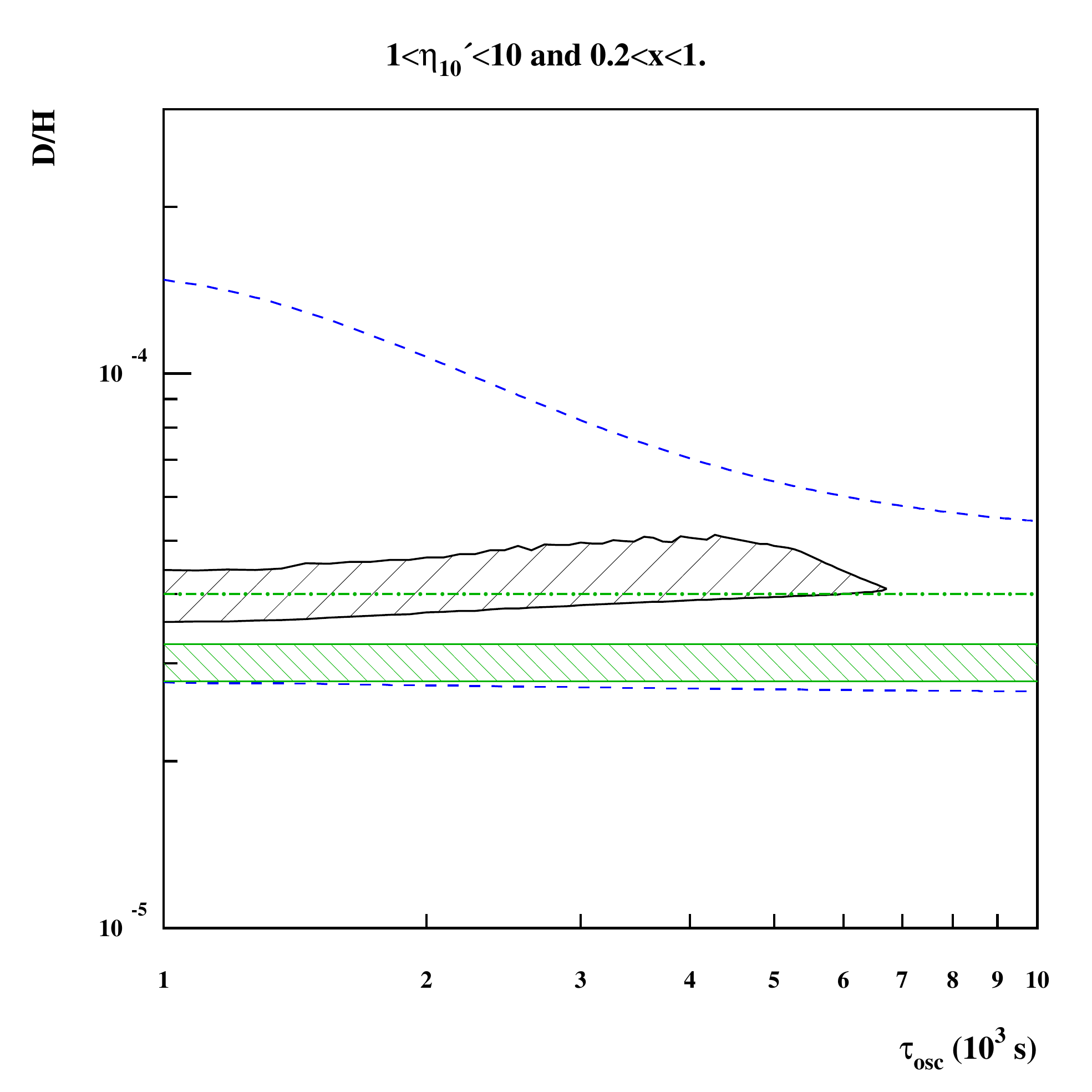}
 \caption{As a function of the oscillation time $\tau_{\rm osc}$, we let $x$ vary between 0.2 and 1 and $\eta'$ between $10^{-10}$ and $10^{-9}$, and explore the value of D/H that can be reached while $\eta=\eta_{\rm WMAP}$ (blue dashed line). Within this range, we find models, i.e. at least one set of parameters $(\eta',x)$ such that D/H, $^7$Li and $Y_p$ would agree with their spectroscopic abundances (black line). This region is a narrow strip around  ${\rm D/H}=4\times10^{-5}$ (dot-dashed line) and exists only if $\tau_{\rm osc}<6\times10^3$~s. The green strip corresponds to ${\rm D/H} =(3.02 \pm 0.23) \times 10^{-5}$ a range for which the model does not allow for any resolution of the lithium-7 problem, whatever the oscillation time.}
 \label{f:global}
\end{figure}

To finish, we let the oscillation time vary and explore the excursion of D/H allowed when $\eta'$ is varying between $10^{-10}$ and $10^{-9}$ and $x$ between $0.2$ and 1, while $\eta$ is fixed to its WMAP value. We then determine the range of values of D/H that would be required in order to have at least 1 model (i.e. a set of $x$ and $\eta'$) so that  D/H, lithium-7 and helium-4  are in agreement with their spectroscopic observations. Figure~\ref{f:global} shows that such a solution exists only if $\tau_{\rm osc}< 6.7\times10^3$~s and would require a typical value of D/H of the order of $4\times10^{-5}$. For larger oscillation times, the mechanism is no more efficient and plays no role on BBN. It is important to note that with the current status on the observation of D/H (i.e. ${\rm D/H} =(3.02 \pm 0.23) \times 10^{-5}$), no global reconciliation is allowed, whatever the value of the oscillation time.

%----------------------------------------------------
%----------------------------------------------------

\section{Conclusions}\label{sec_concl}

This article has investigated the influence of a mirror sector allowing for an oscillation between ordinary and mirror neutrons on the prediction of the abundances of the light elements synthesized during BBN. The main motivation lies in the fact that such a model allows for an injection of neutrons during the late stages of BBN, an effect that in principle shall lead to a decrease of the abundance of lithium-7. It is thus a good candidate in order to resolve the lithium problem, i.e. the fact that the abundance of lithium-7 predicted under the hypothesis that the baryon density is given by the CMB observations is about 3 times larger that the value obtained from the spectroscopic observations.

We have implemented mirror neutrons and neutron oscillation in a BBN code. The model depends on 3 extra-parameters, $x$ the relative mirror photons temperature relative to the CMB temperature today, $\tau_{\rm osc}$ the oscillation time and $\eta'$ the mirror baryon number. 
We have shown that as soon as $x<0.6$ the helium-4 abundance is in agreement with the observations while the agreement for both the deuterium 
and helium-4 abundances can be obtained within a large domain of the parameters space. 
This is a new example of the power of BBN to constrain deviations of the standard model of particle physics.

Concerning the lithium-7 problem, we have shown that if ${\rm D/H} =(3.02 \pm 0.23) \times 10^{-5}$ then  D/H, helium-4 and Li/H cannot be reconciled whatever the parameters $(x,\eta',\tau_{\rm osc})$. However, if ${\rm D/H}\in[3.8,4.6]\times10^{-5}$ then the abundances of all light element (helium-4, deuterium and lithium-7) can be reconciled for a $\eta=\eta_{\rm WMAP}$. In conclusion, there exists a region in the parameter space $(x,\eta',\tau_{\rm osc})$ for which the lithium-7 problem is resolved, at the expense of a moderately higher D/H. Indeed such a value is  not unreasonable, as discussed above. Note that this requires that $\tau_{\rm osc}< 6.7\times10^3$~s. This upper bound on the oscillation time is 15 times larger than the lower bound provided by laboratory experiments so that an other way to falsify this explanation would be to close the window on $\tau_{\rm osc}$.

In conclusion, mirror neutrons do alleviate the lithium-7 problem and neutron oscillation offers an interesting class of falsifiable models that can potentially solve the lithium problem, an issue that can be resolved once spectroscopic measurement of D/H are available or better lower bounds on the oscillation time are established.

\section*{Acknowledgements}
This work was supported by French state funds managed by the ANR within the Investissements d'Avenir programme under reference ANR--11--IDEX--0004--02.
This work was sponsored by the French Agence Nationale pour la Recherche (ANR) via the grant VACOUL (ANR-2010-BLAN-0510-01)
and JPU acknowledges partial support from the ANR via the grant THALES (ANR-10-BLAN-0507-01-02).  
We thank Pasquier Noterdaeme and Patrick Petitjean for fruitful discussions.

%%%%%%%%%%%%%%%%%%%%%%%%%%%%%%%%%%%%%%%%%%%%%%%%%%%


\begin{thebibliography}{99}

\bibitem{jpubook}
 P. Peter and J.-P. Uzan, {\it Primordial cosmology} (Oxford University Press, England, 2009).

\bibitem{dp}
  T.~Damour and B.~Pichon,
  %``Big bang nucleosynthesis and tensor - scalar gravity,''
  Phys.\ Rev.\ D {\bf 59}, 123502 (1999),
  [{\tt astro-ph/9807176}].

\bibitem{bbnus1}
 A.~Coc, K.A.~Olive, J.-P.~Uzan and E.~Vangioni,
  %``Big bang nucleosynthesis constraints on scalar-tensor theories of gravity,''
  Phys.\ Rev.\ D {\bf 73} (2006) 083525,
  [{\tt astro-ph/0601299}].
 
\bibitem{bbnus2}
 A.~Coc, K.~A.~Olive, J.-P.~Uzan and E.~Vangioni,
  %``Non-universal scalar-tensor theories and big bang nucleosynthesis,''
  Phys.\ Rev.\ D {\bf 79}, 103512 (2009)
  [{\tt arXiv:0811.1845}].
 
\bibitem{jpucte}
J.-P. Uzan,
 %``The fundamental constants and their variation: observational status and theoretical motivations'',
 Rev. Mod. Phys. {\bf 75}, 403 (2003), [{\tt hep-ph/0205340}];
 J.-P. Uzan,
 %``Variation of the constants in the late and early Universe'',
 AIP Conf. Proc. {\bf736}, 3 (2004), [{\tt astro-ph/0409424}];
 J.-P. Uzan,
 %``Varying constants, Gravitation and Cosmology'',
 Living Rev. Relat. {\bf4}, 2 (2011), [{\tt arXiv:1009.5514}].

\bibitem{Coc07} 
 A. Coc, N.J. Nunes, K.A. Olive, J.-P. Uzan, and E. Vangioni, 
 %``Coupled Variations of Fundamental Couplings and Primordial Nucleosynthesis'',
 Phys. Rev. D {\bf76}, 023511 (2007), [{\tt astro-ph/0610733}].
 
\bibitem{cdouv}
 A. Coc, P. Descouvemont, K.A. Olive, J.-P. Uzan, and E. Vangioni, 
 %``The variation of fundamental constants and the role of A=5 and A=8 nuclei on primordial nucleosynthesis'',
 Phys. Rev. D {\bf86}, 3529 (2012), [{\tt arXiv:1206.1139}];
 A.~Coc, P.~Descouvemont, J.-P.~Uzan and E.~Vangioni,
 %``Influence of the variation of fundamental constants on the primordial nucleosynthesis,''
 [{\tt arXiv:1301.1474}].

\bibitem{cte-other}  
C.M. M\"uller, G. Sch\"afer,and C. Wetterich, Phys. Rev. D {\bf70}, 083504,
(2004); [{\tt astro-ph/0405373}]; S.J. Landau, M.E. Mosquera, and H.
Vucetich, Astrophys. J. {\bf637}, 38, (2006), [{\tt astro-ph/0411150}]; T.
Dent, S. Stern, and C. Wetterich, Phys. Rev. D {\bf76}, 063513, (2007),
[{\tt arXiv:0705.0696}].

  
\bibitem{keithTASI} 
K.A. Olive, in {\it TASI Lectures on AstroParticle Physics},  [{\tt arXiv:0503065}].

\bibitem{bbnrevue1}
B.~D. Fields,  Annual Review Nuclear and Particle Science {\bf61}, 47 (2011).

\bibitem{bbnrevue2}
F.~Iocco, G.~Mangano,  G.~Miele, O.~Pisanti, and P.~D.~Serpico, Phys. Rep.  {\bf 472}, 1 (2009).

\bibitem{Jed04}
K.~Jedamzik, Phys. Rev. D  {\bf70}, 063524 (2004). 

\bibitem{Alb12}
 D. Albornoz V\'asquez, A. Belikov, A.~Coc, J. Silk and E.~Vangioni 
  Phys. Rev. D \ {\bf87}, 063501 (2012),  [{\tt arXiv:1208.0443}].

\bibitem{LiYang}
  T.D. Lee and C.-N. Yang, Phys. Rev. {\bf 104}, 254 (1956). 
   
\bibitem{mirror-gen}  
  I.Yu. Kobzarev, L.B. Okun, I.Ya. Pomeranchuk, Yad. Fiz. {\bf3}, 1154 (1966);
 S. Blinnikov and M. Khlopov, Sov. Astron. {\bf27}, 371 (1983);
 E. W. Kolb, D. Seckel, and M. S. Turner, Nature {\bf314}, 415 (1985);
 M.Y. Khlopov, G.M. Beskin, N.E. Bochkarev, L.A. Pustylnik, and S.A. Pustylnik, Sov. Astron. {\bf35}, 21 (1991).
 
\bibitem{sterile} 
 R. Foot, H. Lew, and R. R. Volkas, Phys. Lett. B {\bf272}, 67 (1991);
 E.K. Akhmedov, Z.G. Berezhiani, and G. Senjanovi\' c, Phys. Rev. Lett. {\bf69}, 3013 (1992).

\bibitem{mirror-revue}
 P. Ciarcelluti, Int. J. Mod. Phys. D {\bf19}, 2151 (2010);
 L. B. Okun, Phys. Usp.  {\bf50}, 380 (2006), [{\tt hep-ph/0606202}];
 R. Foot, Int. J. Mod. Phys. A {\bf19}, 3807 (203), [{\tt astro-ph/0309330}].
 
\bibitem{mirror-osc}
 E.D. Carlson, and S.L. Glashow, Phys. Lett. B {\bf193}, 168 (1987);
 R. Foot, and R.R. Volkas, Phys. Rev. D {\bf52}, 6595 (1995);
 Z. Berezhiani, and R.N. Mohapatra, Phys. Rev. D {\bf 52}, 6607 (1995).
   
\bibitem{mirrorneutron}
 Z. Berezhiani, and L. Bento, Phys. Rev. Lett. {\bf96}, 081801 (2006); 
 Z. Berezhiani, and L. Bento, Phys. Lett. B {\bf 635}, 253 (2006).
 
\bibitem{mirrorDM} 
 Z. Berezhiani, Int. J. Mod. Phys. A {\bf19}, 3775 (2004), [{\tt hep-ph/0508233}];
 Z. Berezhiani, and R. Foot, Int. J. Mod. Phys. A {\bf19}, 3807 (2004), [{\tt astro-ph/0309330}];
 S. Blinnikov, and M. Khlopov, Sov. Astron. {\bf 27}, 371 (1983);
 Z. Berezhiani, A.D. Dolgov, and R.N. Mohapatra, Phys. Lett. B {\bf375}, 26 (1996);
 L. Bento, and Z. Berezhiani, Phys. Rev. Lett. {\bf87}, 231304 (2001), [{\tt hep-ph/0111116}]; 
 A.Y. Ignatiev, and R.R. Volkas, Phys. Rev. D {\bf68}, 023518 (2003);
 Z. Berezhiani, \etal, Int. J. Mod. Phys. D {\bf14}, 107 (2005).

\bibitem{mirrorBBNearly}
 Z. Berezhiani,  D. Comelli, and F.L. Villante,  Phys. Lett. B {\bf503}, 362 (2001).
 
\bibitem{DMnew}
 R. Foot, Phys. Rev. D {\bf78}, 043529 (2008), [{\tt arXiv:0804.4518}],
 R. Foot, Phys. Rev. D {\bf86}, 023524 (2012), [{\tt arXiv:1203.2387}];
 R. Foot, Phys. Lett. B {\bf711}, 238 (2012);
 R. Foot, [{\tt arXiv:1211.3217}];
 P. Ciarcelluti, and Q. Wallemacq, [{\tt arXiv:1211.5354}].

\bibitem{mirrorlab1}
 G. Ban, \etal, Phys. Rev. Lett. {\bf99}, 161603 (2007), [{\tt arXiv:0705.2336}].
 
\bibitem{mirrorlab2} 
 A.P. Serebrov, \etal, Phys. Lett. B {\bf663}, 181 (2008), [{\tt arXiv:0706.3600}];
 
\bibitem{exp-mirror-n1} 
 A.P. Serebrov, E.B. Aleksandrov, N.A. Dovator, S.P. Dmitriev, \etal,
 Nuclear Instruments and Methods in Physics Research A {\bf 611}, 137 (2009).
 
\bibitem{PDG}
The Review of Particle Physics, 
J. Beringer \etal, (Particle Data Group), Phys. Rev. D {\bf86}, 010001 (2012).  

\bibitem{groom}
 D. Groom, \etal, Eur. Phys. J. C {\bf15}, 1 (2000).

\bibitem{bbnNeff}
 G. Mangano, and P.D. Serpico, 
 %ÒA robust upper limit on Neff from BBN, circa 2011Ó, 
 Phys. Lett. B {\bf701}, 296 (2011).
  
\bibitem{spitex2} 
 F. Spite, and M. Spite, Astron. Astrophys. {\bf 115}, 357 (1982).
 
\bibitem{bm} 
 P. Bonifacio, and P. Molaro, Month. Not. R. Astron. Soc. {\bf285}, 847 (1997).

\bibitem{rnb} 
 S.G. Ryan, J. Norris, and T.C.  Beers, Astrophys. J. {\bf523}, 654 (1999).
 
\bibitem{Asplundetal06} 
 M. Asplund, D.L. Lambert, P.E. Nissen, F. Primas, and V.V. Smith, Astrophys. J. {\bf644}, 229 (2006).

\bibitem{Bonifacioetal07} 
 P. Bonifacio, P. Molaro, P., T. Sivarani, R. Cayrel, R., \etal, Astron. Astrophys. {\bf462}, 851 (2007).
 
\bibitem{hos} 
 A. Hosford, S.G. Ryan, A.E. Garc{\'{\i}}a P{\'e}rez, and K.A. Olive,  Astron. Astrophys. {\bf493}, 601 (2009).

\bibitem{hos2} 
 A. Hosford, A.E. Garc{\'{\i}}a P{\'e}rez, R. Collet, S.G. Ryan, \etal, Astron. Astrophys. {\bf511}, 47 (2010).

\bibitem{sbordone} 
 L. Sbordone, P. Bonifacio, E. Caffau, \etal, Astron. Astrophys. {\bf522}, A26 (2010).
 
\bibitem{ss4} 
  F. Spite, and M. Spite, IAU Symposium proceedings, {\bf268}, Charbonnel \etal~Eds.
  (Cambridge Univ. Press, 2010) p. 201.

\bibitem{WMAP11} 
 E. Komatsu, K.M. Smith, J. Dunkley, \etal,
 Astrophys. J. Suppl. Series {\bf192}, 18 (2011).

\bibitem{coc12} 
 A. Coc, S. Goriely, Y. Xu, M. Saimpert, and E. Vangioni, 
 Astrophys. J. {\bf744}, 158 (2011).
 
\bibitem{Ryan00} 
 S.G. Ryan, T.C.  Beers, K.A. Olive, B.D. Fields, and J.E.  Norris, Astrophys. J. {\bf530}, L57 (2000).

\bibitem {Cocetal04}
 A. Coc, E. Vangioni-Flam, P. Descouvemont, A. Adahchour, and C. Angulo,  Astrophys. J. {\bf600}, 544 (2004).

\bibitem{angulo} 
 C. Angulo, E. Casarejos, M. Couder, \etal, Astrophys. J.  {\bf630}, L105  (2005).
 
\bibitem{cfo4}  
 R.H. Cyburt, B.D. Fields,  and K.A. Olive, Phys. Rev. D {\bf69}, 123519 (2004).

\bibitem{nuclear} 
R.H.~Cyburt \& M.~Pospelov, 
{\it International Journal of Modern Physics} {\bf E21}, 50004 (2012);
P.D.~O'Malley, D.W.~Bardayan, A.S.~Adekola \etal, 
Phys. Rev. C  {\bf 84}, 042801 (2011);
O.S.~Kirsebom \& B.~Davids,
Phys. Rev. C  {\bf 84}, 058801 (2011);
C.~Scholl, Y.~Fujita, T.~Adachi \etal,  Phys. Rev. C   {\bf 84}, 014308 (2011);
N.~Chakraborty, B.D.~Fields \& K.A.~Olive, Phys. Rev. D  {\bf 83}, 063006 (2011);
C.~Broggini, L.~Canton, G.~Fiorentini \& F.L.~Villante,
J. Cosmol. Astropart. Phys.  {\bf 6}, 30 (2012);
O.~Civitarese \& M.E.~Mosquera, Nucl. Phys. A {\bf 898}, 1 (2013).


 \bibitem{boyd} 
  R.N. Boyd, C.R. Brune, G.M. Fuller, and C.J. Smith, phys. Rev. D {\bf82}, 105005 (2010).
  
\bibitem{Ric05}
O.~Richard, G.~Michaud, \& J.~Richer, Astrophys. J. {\bf 619}, 538 (2005). 

  
\bibitem{Kornetal06} 
 J. Korn, F. Grundahl, O. Richard, P.S. Barklem, \etal, Nature {\bf442}, 657 (2006).
 
\bibitem{GarciaPerezetal08} 
 A.E. Garc\'{i}a Per\'{e}z, S. Inoue, W. Aoki, and S.G. Ryan,
 Precision Spectroscopy in Astrophysics, Proceedings of the ESO/Lisbon/Aveiro Conference, 9 (2008).
 
 \bibitem{mr2} 
 J. Mel\'{e}ndez, L. Casagrande, I. Ram\'{i}rez, M. Asplund, and W.J.  Schuster, Astron. Astrophys. {\bf515}, 3  (2010).
 
 \bibitem{Cay07}
 R.~Cayrel, M. Steffen, H.~Chand, P.~Bonifacio, M.~Spite, F.~Spite4,  P.~Petitjean, H.-G.~Ludwig \& E.~Caffau,
 Astron. Astrophys. {\bf 473} L37 (2007).
 
\bibitem{eov} 
 J. Ellis, K.A.  Olive, and E. Vangioni, Phys. Lett. B {\bf 619}, 30 (2005).
 
\bibitem{Jedamzik06} 
 K. Jedamzik, K.-Y. Choi, L. Roszkowski, and R. Ruiz de Austri, JCAP {\bf 0607}, 7 (2006). 
 
\bibitem{pps} 
 M. Pospelov, J. Pradler, and Steffen, JCAP {\bf0811}, 20 (2008).
 
\bibitem{jp} 
 K. Jedamzik, and M.  Pospelov, New J. Phys. {\bf11}, 105028  (2009).
 
\bibitem{grant3} 
 M. Kusakabe, T. Kajino, T. Yoshida,  and G.J. Mathews, Phys. Rev. D {\bf81}, 083521 (2010).

\bibitem{Cyb13}
R.H.~Cyburt, J.~Ellis, B.D.~Fields, F.~Luo, K.A.~Olive, V.C.~Spanos, 
[{\tt arXiv:1303.0574}].

\bibitem{pp2} 
 M. Pospelov, and J. Pradler, Phys. Rev. D {\bf82}, 103514 (2010). 
 
\bibitem{jed08b}
K. Jedamzik,   Phys. Rev. D {\bf 77}, 063524 (2008).
 
\bibitem{ceflos1.5} 
 R.H. Cyburt, J. Ellis,  B.D. Fields, \etal, JCAP {\bf1010}, 32 (2010).

\bibitem{jittoh2} 
 T. Jittoh, K. Kohri, M. Koike, \etal, Phys. Rev. D {\bf82}, 115030 (2010).

\bibitem{kk} 
 M. Kawasaki, and M.  Kusakabe, Phys. Rev. D {\bf83}, 055011 (2011).

\bibitem{sik} 
 O. Erken, P. Sikivie, H. Tam, and Q. Yang, Phys. Rev. Lett. {\bf108}, 061304  (2012). 

\bibitem{kus} 
 M. Kusakabe, A.B. Balantekin, T. Kajino, and Pehlivan, Phys. Lett. B {\bf 718} 704 (2013) [{\tt arXiv:1202.5603}].  

\bibitem{dfw} 
 V.F. Dmitriev, V.V. Flambaum, and J.K. Webb, Phys. Rev. D {\bf69}, 063506 (2004).

\bibitem{cp1} 
  M.~Regis and C.~Clarkson,
  %``Do primordial Lithium abundances imply there's no Dark Energy?,''
  Gen.\ Rel.\ Grav.\  {\bf 44}, 567 (2012)
  [{\tt arXiv:1003.1043}].
  
\bibitem{cp2}  
 J.~-P.~Uzan,
  %``Dark energy, gravitation and the Copernican principle,''
  [{\tt arXiv:0912.5452}].
  
\bibitem{LiinC}
 F. Iocco, P. Bonifacio, and E. Vangioni, Proceedings of the workshop {\em Lithium in the cosmos},
 Mem. Soc. Aastron. It. Suppl. {\bf 22}, 3 (2012). 

\bibitem{wmap9}
 G. Hinshaw, D. Larson, E. Komatsu, \etal, [{\tt arXiv:1212.5226}].

\bibitem{os1}  
 K.A. Olive, and E.D. Skillman, New Astronomy {\bf6}, 119 (2001).

\bibitem{aos3} 
 E. Aver, K.A. Olive, and E.D. Skillman, JCAP {\bf1204}, 004 (2012).

\bibitem{Olive2012} 
  K.A. Olive, P. Petitjean, E. Vangioni, and J. Silk, 
 Month. Not. R. Astron. Soc. {\bf 426}, 1427 (2012).
 
\bibitem{pettini2} 
 M. Pettini, B.J. Zych, M.T. Murphy, A. Lewis, and C.C. Steidel, Month. Not. R. Astron. Soc. {\bf391}, 1499 (2008).
 
\bibitem{fuma} 
 M. Fumagalli, J.M. O'Meara, and J.X. Prochaska, [{\tt arXiv:1111.2334}].  
 
\bibitem{pettini12} 
 M. Pettini, and R. Cooke, Month. Not. R. Astron. Soc. {\bf425}, 2477 (2012).

\bibitem{Des04}
 P.~Descouvemont, A.~Adahchour, C.~Angulo, A.~Coc A, and E.~Vangioni--Flam,
 {\it Atomic Data and Nuclear Data Tables} {\bf 88}, 203 (2004).

\bibitem{And06}
 S. Ando, R.H. Cyburt, S.W. Hong, and C.H. Hyun,
 Phys. Rev. C {\bf74}, 025809 (2006).

\bibitem{CV10}
A.~Coc \& E.~Vangioni, {\it Journal of Physics Conference Series} {\bf 202}, 012001 (2010).

 \bibitem{Cyb08a} 
 R.H. Cyburt and B.~Davids,
 Phys. Rev. C {\bf 78}, 064614 (2008).
 
\bibitem{ENAS6}
A.~Coc, {\em Proceedings of the VI European Summer School on Experimental Nuclear Astrophysics},
PoS(ENAS 6)017, Published online at {\tt http://pos.sissa.it/cgi-bin/reader/conf.cgi? confid=148"}.

\bibitem{Moh80}    
R.N. Mohapatra and R.E. Marshak, Phys. Rev. Lett. B \ {\bf 94}, 183 (1980).

\bibitem{vangioni88} 
 E. Vangioni-Flam, and J. Audouze, Astron. Astrophys. {\bf193}, 81 (1988).


\end{thebibliography}
\end{document}